\documentclass[11pt]{article}
\usepackage{amsmath,amssymb,amsthm,amsxtra,overpic,bbm,bm,epsfig,subfigure}
\usepackage{color}
\usepackage{comment}
\usepackage{cancel}
\usepackage{multirow}

\def\thefootnote{\fnsymbol{footnote}}

\begin{document}

\vspace{0.2cm}

\begin{center}
{\Large\bf Improved Statistical Determination of Absolute Neutrino Masses via Radiative Emission of Neutrino Pairs from Atoms}
\end{center}

\vspace{0.2cm}

\begin{center}
{\bf Jue Zhang $^{a}$} \footnote{E-mail: zhangjue@ihep.ac.cn}
\quad {\bf Shun Zhou $^{a,~b}$} \footnote{E-mail: zhoush@ihep.ac.cn}
\\
{$^a$Institute of High Energy Physics, Chinese Academy of
Sciences, Beijing 100049, China \\
$^b$Center for High Energy Physics, Peking University, Beijing 100871, China}
\end{center}

\vspace{1.5cm}
\begin{abstract}
The atomic transition from an excited state $|{\rm e}\rangle$ to the ground state $|{\rm g}\rangle$ by emitting a neutrino pair and a photon, i.e., $|{\rm e}\rangle \to |{\rm g}\rangle + |\gamma\rangle + |\nu^{}_i\rangle + |\overline{\nu}^{}_j\rangle$ with $i, j = 1, 2, 3$, has been proposed by Yoshimura and his collaborators as an alternative way to determine the absolute scale $m^{}_0$ of neutrino masses. More recently, a statistical analysis of the fine structure of the photon spectrum from this atomic process has been performed [N. Song {\it et al.}, Phys.\ Rev.\ D {\bf 93}, 013020 (2016)] to quantitatively examine the experimental requirements for a realistic determination of absolute neutrino masses. In this paper, we show how to improve the statistical analysis and demonstrate that the previously required detection time can be reduced by one order of magnitude for the case of a $3\sigma$ determination of $m_0^{} \sim 0.01~\text{eV}$ with an accuracy better than $10\%$. Such an improvement is very encouraging for further investigations on measuring absolute neutrino masses through atomic processes.
\end{abstract}

\begin{flushleft}
\hspace{0.8cm} PACS number(s): 14.60.Pq, 32.80.Qk
\end{flushleft}

\def\thefootnote{\arabic{footnote}}
\setcounter{footnote}{0}

\newpage

\section{Introduction}

Dedicated experimental efforts in the last two decades have greatly improved our knowledge on neutrinos. It is now a well-established fact that neutrinos are massive and significantly mixed among three flavors~\cite{Agashe:2014kda}. The ongoing and forthcoming neutrino experiments will further unravel the mysteries of neutrinos, such as the neutrino mass ordering, the size of CP violation in the lepton sector, the absolute scale of neutrino masses and the Majorana or Dirac nature of neutrinos (i.e., whether neutrinos are their own antiparticles).

Concerning the determination of the absolute scale $m_0^{}$ of neutrino masses, there currently exist three major experimental approaches. First, one can study the kinematic impact of massive neutrinos on the electron energy spectrum near the end point from nuclear beta decays, such as the KATRIN experiment~\cite{Osipowicz:2001sq}. Second, one can measure the half-lives of the neutrinoless double-beta decays ($0\nu\beta\beta$) of some even-even heavy nuclei and extract the effective neutrino mass, which is related to absolute neutrino masses and other flavor mixing parameters (see, e.g., Ref.~\cite{Pas:2015eia} for a recent review). Finally, the information about absolute neutrino masses can be inferred from the power spectrum of cosmic microwave background and the large-scale structure of the Universe~\cite{Wong:2011ip}. According to the latest Planck data~\cite{Ade:2015xua}, we now know the upper bound on the sum of all three neutrino masses $m^{}_1 + m^{}_2 + m^{}_3 < 0.23~\text{eV}$ at the $95\%$ confidence level.

As the characteristic energy scale in atomic physics is at the eV level, atomic processes should be an ideal place to measure neutrino masses. In a series of papers in the past few years~\cite{RENP_idea,Fukumi:2012rn,Dinh:2012qb}, Yoshimura and his collaborators (also known as the SPAN group) have proposed to study a special atomic transition process $|{\rm e}\rangle \rightarrow |{\rm g}\rangle + |\gamma\rangle + |\nu_i^{}\rangle + |\overline{\nu}_j^{}\rangle$, where $|{\rm e}\rangle$ and $|{\rm g}\rangle$ are the excited and ground energy levels of atoms, $|\gamma\rangle$ denotes the emitted photon, and the associated radiative emission of a neutrino pair (RENP) is indicated by $|\nu_i^{}\rangle + |\overline{\nu}_j^{}\rangle$, with $|\nu_i^{}\rangle$ (for $i = 1, 2, 3$) being neutrino mass eigenstates. Since the energy difference between $|{\rm e}\rangle$ and $|{\rm g}\rangle$ is known to a very good accuracy, one can determine absolute neutrino masses through a precise measurement of the energy spectrum of the emitted photon. Recently, a statistical analysis of the photon spectrum from the RENP process has been carried out in Ref.~\cite{Song:2015xaa}, where the experimental requirements for a statistical determination of neutrino mass scale are quantitatively investigated. It has been found that even under ideal conditions, such as a perfect coherence among target atoms and an excellent detection efficiency, determining neutrino masses with a good precision requires a detection time on the order of a few days to months, which seems to be beyond the reach of current technologies. For instance, it has been shown that a few months are needed for a $3\sigma$ determination of $m_0^{} \sim 0.01~\text{eV}$ with an accuracy better than $10\%$~\cite{Song:2015xaa}. In this paper, we improve the previous work by a detailed analysis of the photon spectrum, and demonstrate that one can reduce the detection time by one order of magnitude by looking into the second kink in the fine structure. Although such a reduction may be still inadequate for building a successful RENP experiment at present, our finding is very encouraging for future investigations in this direction.

The remaining part is organized as follows. In Section 2 we briefly review the general idea of the RENP proposal. An analytical analysis of the spectrum function of the emitted photon is given in Section 3. In Section 4 we present our main results of determining neutrino masses. Finally, we conclude in Section 5.

\section{Radiative Emission of Neutrino Pairs}
%%%%%%%%%%%%%%%%%%%%%%%% Fig. 1 %%%%%%%%%%%%%%%%%%%%%%%%%%%%%%%%%%%%%%%%%%%%
\begin{figure}[t]
\centering
\includegraphics[scale=0.6]{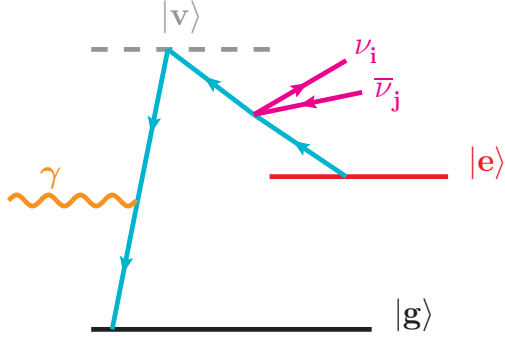}
\caption{A schematic diagram for radiative emission of a neutrino pair in a three-level atomic system, where $|{\rm e}\rangle$, $|{\rm v}\rangle$ and $|{\rm g}\rangle$ denote the excited, intermediate virtual and ground states of atoms, respectively.}
\label{fg:RENP_Lambda}
\end{figure}
%%%%%%%%%%%%%%%%%%%%%%%%%%%%%%%%%%%%%%%%%%%%%%%%%%%%%%%%%%%%%%%%%%%%%%%%%%%%

The physical picture of the RENP process is a two-step successive transition in a three-level system, as shown in Fig.~\ref{fg:RENP_Lambda}, where the atomic system consists of an excited state $|{\rm e}\rangle$, an intermediate virtual state $|{\rm v}\rangle$ and a ground state $|{\rm g}\rangle$. For the occurrence of RENP, several conditions for the atomic system need to be satisfied~\cite{Fukumi:2012rn}. First, the direct electric dipole (E1) transition between the excited state and the ground state is forbidden. Otherwise, the excited state would quickly decay into the ground state via photon emission. Second, the transition between the excited state and the virtual state is also E1 forbidden but magnetic dipole (M1) allowed. It can be shown that the two-level atomic transition involving neutrino pair emission is M1-like~\cite{Fukumi:2012rn}, so we reserve this transition for RENP. Of course, the photon emission via M1 transition is also allowed, and it may constitute as an important background for the RENP observation. Lastly, we allow the E1 transition from the virtual state to the ground state so that photon emission is dominant in such a transition. In summary, by choosing a special atomic state configuration we can observe RENP in association with a photon emission during such a two-step atomic transition.

Although a real RENP experiment has not yet been achieved so far, the basic idea to measure the photon energy spectrum seems to be reasonable~\cite{RENP_idea,Fukumi:2012rn}. First of all, a population inversion between the ground and excited states should be realized. Moreover, as the RENP process involves a high-order and weak interaction, it is crucial to maintain a remarkable coherence among atoms such that the effect of multiatomic coherent emission~\cite{super_radiance} can be implemented to significantly enhance the transition rate. Next, to stimulate the RENP process, one irradiates the target atoms by two beams of counterpropagating trigger lasers, whose frequencies $\omega_1^{}$ and $\omega_2^{}$ satisfy $\omega_1^{} < \omega_2^{}$ and $\omega_1^{} + \omega_2^{} = \epsilon_{\rm eg}^{}$, where $\epsilon_{\rm x y}^{} \equiv \epsilon^{}_{\rm x} - \epsilon^{}_{\rm y}$ is the difference between the energy $\epsilon^{}_{\rm x}$ of an atomic state $|{\rm x}\rangle$ and that $\epsilon^{}_{\rm y}$ of another state $|{\rm y}\rangle$. Because of the energy-momentum conservation, the photon is then emitted in the direction of the laser with a lower frequency, i.e., $\omega_1^{}$. Finally, we measure the intensity of photons emitted during such a trigger laser irradiation. By varying the frequency $\omega_1^{}$ of the trigger laser ($\omega_2^{}$ will also be changed accordingly), we obtain the energy spectrum of the emitted photons, from which valuable information on neutrino properties can be extracted.

The formula of the RENP rate in the three-level atomic system has been given in Refs.~\cite{Fukumi:2012rn,Dinh:2012qb,Song:2015xaa}, where the details of derivation can be found. Defining the RENP rate as the number of emitted photons of a frequency $\omega$ per unit time, we obtain~\cite{Fukumi:2012rn,Dinh:2012qb,Song:2015xaa}
\begin{eqnarray}
\frac{{\rm d} N_\gamma^{} (\omega) }{{\rm d} t} = (0.481~{\rm Hz}) (2J_{\rm v}^{} + 1) C_{\rm ev} \left( \frac{V}{10^2_{}~{\rm cm}^3_{}} \right ) \left( \frac{n}{10^{21}_{}~{\rm cm}^{-3}_{}} \right) \left( \frac{\gamma_{\rm vg}^{}}{10^8_{} ~{\rm Hz}} \right) \left( \frac{\epsilon_{\rm eg}^{}}{{\rm eV}} \right) \left( \frac{{\rm eV}}{\epsilon_{\rm vg}^{}} \right)^3_{} \mathcal{I}(\omega) ~ \eta_\omega^{}(t) \; ,
\end{eqnarray}
where $J_{\rm v}^{}$ is the angular-momentum quantum number of the atomic state $|{\rm v}\rangle$, $C_{\rm ev}$ is the atomic spin factor whose definition can also be found in Ref.~\cite{Song:2015xaa}, $V$ is the target volume, $n$ is the number density of atoms in the target, and $\gamma_{\rm vg}^{}$ is the spontaneous dipole transition rate between states $|{\rm v}\rangle$ and $|{\rm g}\rangle$. Lastly, $\mathcal{I}(\omega)$ and $\eta_\omega^{}(t)$ are the photon energy spectrum function and the medium dynamical factor, respectively. As an example, in this work we focus on the atomic candidate ${\rm Yb}$, whose relevant energy levels and atomic parameters have been identified in Refs.~\cite{Fukumi:2012rn,Dinh:2012qb,Song:2015xaa} and are now summarized as follows:
\begin{eqnarray} \label{eq:atom_paras}
\epsilon_{\rm eg}^{} = 2.14349~\text{eV}, \quad
\epsilon_{\rm vg}^{} = 2.23072~\text{eV}, \quad
\gamma_{\rm vg}^{} = 0.0115\times 10^8_{}~\text{Hz}, \quad
(2J_{\rm v}^{}+1)C_{\rm ev} = 2 \; .
\end{eqnarray}
The precise definition of $\eta_\omega^{}(t)$ can also be found in Refs.~\cite{Fukumi:2012rn,Dinh:2012qb,Song:2015xaa}, and it characterizes the level of coherence among atoms in the target. In a recent series of experiments \cite{PSR_exps_SPAN} searching for paired superradiance, a twin process of RENP, the SPAN group has reported an induced coherence of about $6.5\%$ using the method of adiabatic Raman scattering. As reaching a high level of coherence is crucial to enhance the RENP rate, more efforts have recently been devoted to exploring various experimental techniques from quantum optics. For instance, the authors of Ref.~\cite{Boyero:2015eqa} suggest the technique of coherent population return, and claim that a coherence close to $100\%$ might be achievable. For simplicity, in our phenomenological study we assume $\eta_\omega^{}(t) = 100\%$, namely, a full coherence among all atoms.

The spectrum function $\mathcal{I}(\omega)$ of emitted photons carries valuable information of neutrino properties, and its analytical expression turns out to be~\cite{Fukumi:2012rn,Dinh:2012qb,Song:2015xaa}
\begin{eqnarray}
\text{Dirac}: & &\mathcal{I}(\omega) = \frac{1}{(\omega - \epsilon_{\rm vg}^{})^2} \sum_{ij}^{} \Delta_{ij}^{} (\omega)  |a_{ij}^{}|^2 I_{ij}^{}(\omega)~ \Theta( \omega_{ij}^{} - \omega ) \; , \\
\text{Majorana}: & &\mathcal{I}(\omega) = \frac{1}{(\omega - \epsilon_{\rm vg}^{})^2} \sum_{ij}^{} \Delta_{ij}^{} (\omega)  \left[ |a_{ij}^{}|^2 I_{ij}^{}(\omega) - m_i^{} m_j^{} \text{Re}(a_{ij}^2) \right ]~ \Theta( \omega_{ij}^{} - \omega ) \; ,
\end{eqnarray}
with $\Delta_{ij}(\omega)$ and $I_{ij}^{}(\omega)$ defined as
\begin{eqnarray}
\Delta_{ij}(\omega) &\equiv& \frac{ \left[ \epsilon_{\rm eg}^{} ( \epsilon_{\rm eg}^{} - 2\omega ) - (m_i^{} + m_j^{})^2 \right]^{1/2}_{} \left[ \epsilon_{\rm eg}^{} ( \epsilon_{\rm eg}^{} - 2 \omega ) - (m_i^{} - m_j^{})^2 \right]^{1/2}_{}}{\epsilon_{\rm eg}^{} ( \epsilon_{\rm eg}^{} - 2 \omega )} \; , \\
I_{ij}^{}(\omega) &\equiv& \frac{1}{3} \left[ \epsilon_{\rm eg}^{}(\epsilon_{\rm eg}^{} - 2\omega) + \frac{\omega^2}{2} - \frac{\omega^2 \Delta_{ij}^2(\omega)}{6} - \frac{(m_i^2 + m_j^2)}{2} - \frac{(\epsilon_{\rm eg}^{} - \omega)^2 (m_i^2 - m_j^2)}{2\epsilon_{\rm eg}^2(\epsilon_{\rm eg}^{} - 2 \omega)^2} \right ] \; .
\end{eqnarray}
In the above equations, we have distinguished the cases of Dirac and Majorana neutrinos, as in the latter case a term proportional to neutrino masses appears. In addition, $m_i^{}$ (for $i = 1, 2, 3$) are neutrino masses,  $a_{ij}^{} \equiv U_{ei}^* U_{ej}^{} - \delta_{ij}^{}/2$ with $U_{ei}^{}$ being the $i$th element in the first row of the lepton flavor mixing matrix, and $\Theta(\omega_{ij}^{} -\omega)$ denotes the Heaviside function with the six threshold locations $\omega^{}_{ij}$ given by~\cite{RENP_idea}
\begin{eqnarray}\label{eq:threshold}
\omega_{ij}^{} = \frac{\epsilon_{\rm eg}^{}}{2} - \frac{(m_i^{} + m_j^{})^2}{2\epsilon_{\rm eg}^{}} \; .
\end{eqnarray}
The fine structure caused by nonzero neutrino masses in the photon energy spectrum can be explored by scanning over a certain range of trigger laser frequencies.

Since the spectrum functions in Eqs.~(3) and (4) depend on almost all the mass and mixing parameters of neutrinos, it has been suggested in Ref.~\cite{Dinh:2012qb} that a precise determination of the photon spectrum can answer almost all currently unknown questions about neutrinos. However, not all of these unknown issues will be addressed with the same sensitivity in a real RENP experiment. In the order of increasing difficulty of their determinations, one may roughly rank them as: the absolute neutrino masses, the neutrino mass ordering, the Dirac or Majorana nature of neutrinos and the Majorana CP-violating phases. Since the last two goals are not expected to be achieved in the near future and the neutrino mass ordering is likely to be determined within ten years by neutrino oscillation experiments \cite{mass_order_neu_exps}, the primary aim of RENP experiments is then to pin down the absolute scale of neutrino masses in the immediate future. Therefore, in this work we are going to focus only on the issue of measuring neutrino masses and consider the case of massive Dirac neutrinos for illustration. Certainly, the discrimination between Dirac and Majorana neutrinos deserves a further and separate study.

\section{Fine Structure of Photon Spectrum}
%%%%%%%%%%%%%%%%%%%%%%%%%%%%%%%%%%%% Fig. 2 %%%%%%%%%%%%%%%%%%%%%%%%%
\begin{figure}[t]
\centering
\includegraphics[scale=0.65]{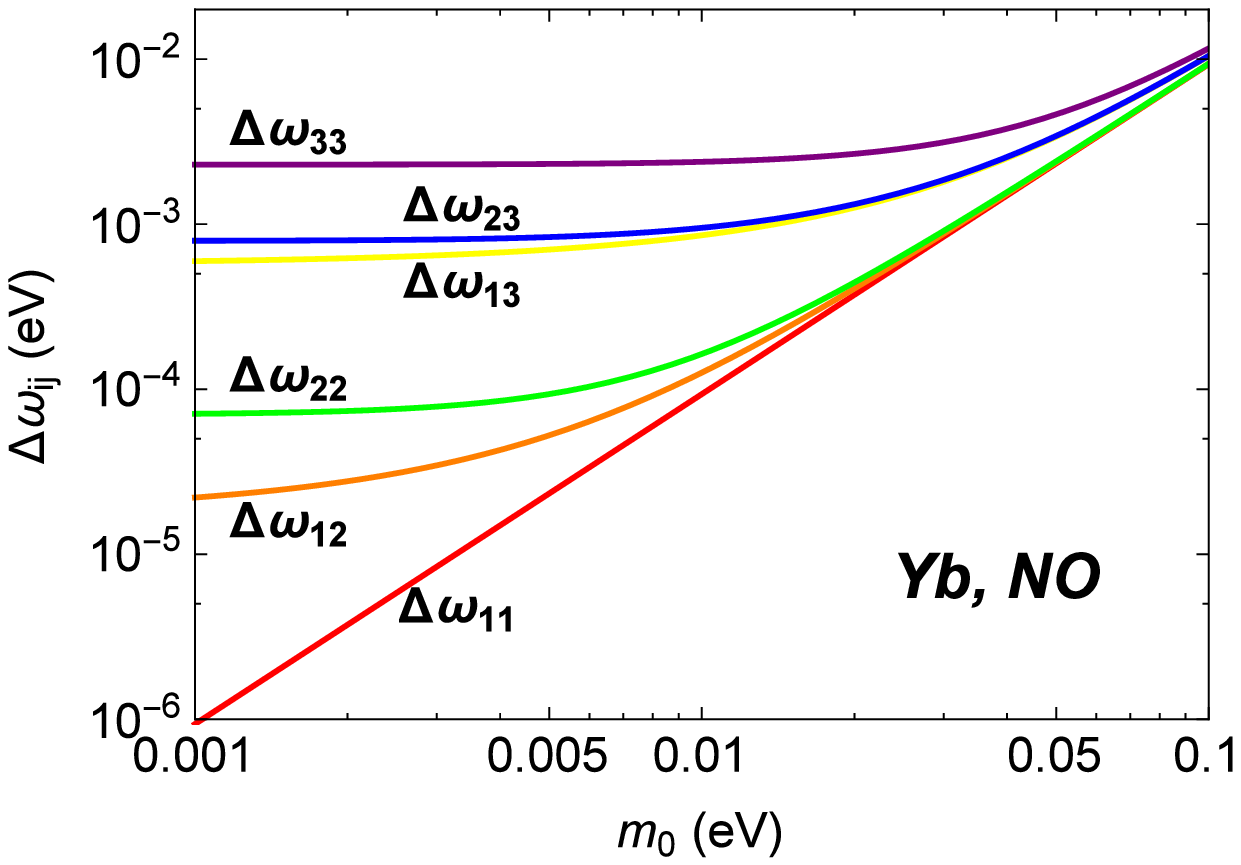}~
\includegraphics[scale=0.65]{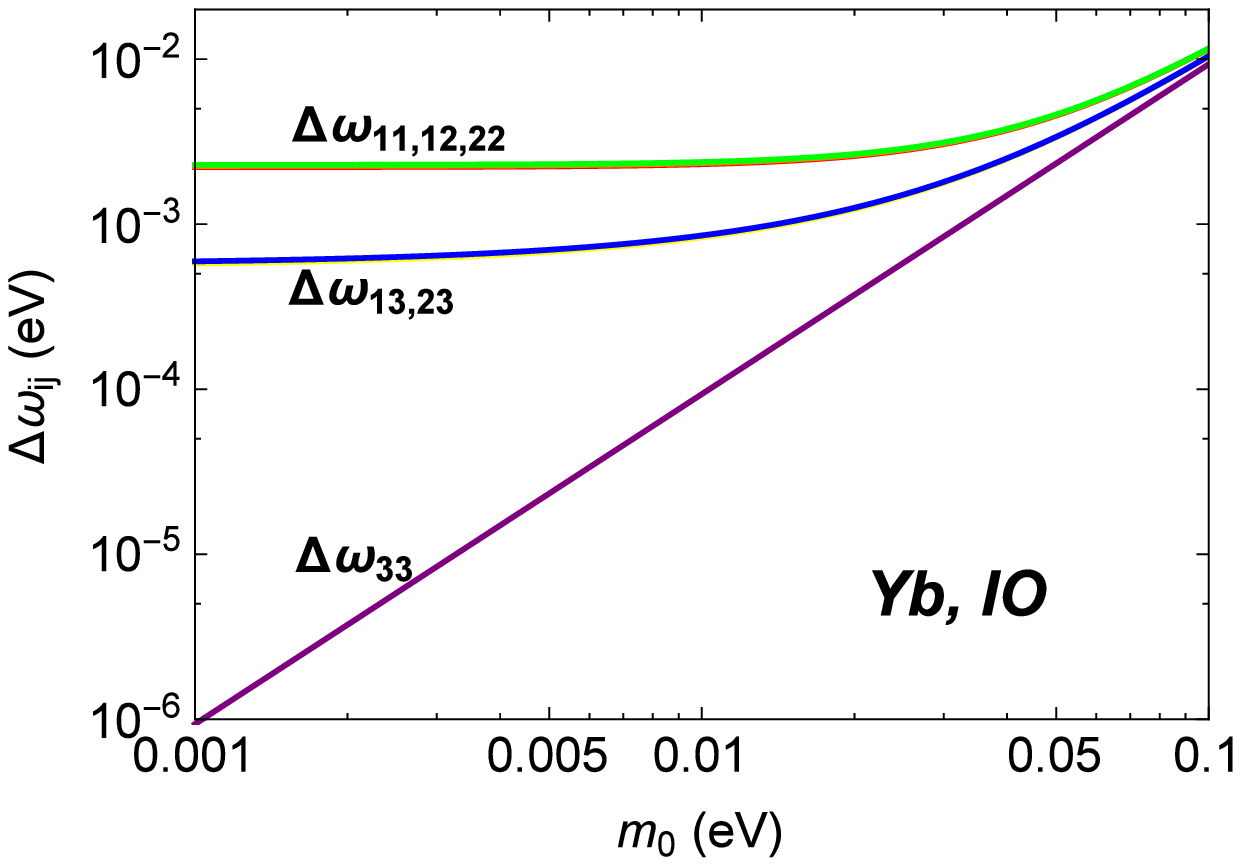}
\caption{Illustration for the deviation $\Delta\omega_{ij}^{} \equiv \epsilon^{}_{\rm eg}/2 - \omega^{}_{ij} = (m^{}_i + m^{}_j)^2/(2\epsilon^{}_{\rm eg})$ as a function of the lightest neutrino mass $m_0^{} = m^{}_1$ for NO (left) and $m^{}_0 = m^{}_3$ for IO (right), where $\epsilon^{}_{\rm eg} = 2.14349~{\rm eV}$, $\Delta m_{21}^2 = 7.50 \times 10^{-5}_{}~\text{eV}^2$ and $\Delta m_{31}^2 = 2.457 \times 10^{-3}_{}~\text{eV}^2$ for NO ($\Delta m_{23}^2 = 2.449 \times 10^{-3}_{}~\text{eV}^2$ for IO) have been used.}
\label{fg:dev}
\end{figure}
%%%%%%%%%%%%%%%%%%%%%%%%%%%%%%%%%%%%%%%%%%%%%%%%%%%%%%%%%%%%%%%%%%%%%
In this section we provide a detailed discussion on the fine structure of the spectrum function $\mathcal{I}(\omega)$ in Eq.~(3). Let us start with the numerical factors $|a_{ij}^{}|^2$ and the threshold locations $\omega_{ij}^{}$, given the current knowledge of neutrino masses and mixing parameters from neutrino oscillation data. First of all, the unitarity of the lepton flavor mixing matrix leads to a useful identity $\sum_{i,j} |a_{ij}^{}|^2 = 3/4$. Adopting the best-fit values of three lepton mixing angles from Ref.~\cite{Gonzalez-Garcia:2014bfa} (namely, $\theta_{12}^{} = 33.5^\circ$ and $\theta_{13}^{} = 8.5^\circ$) for both normal neutrino mass ordering (NO), i.e., $m^{}_1 < m^{}_2 < m^{}_3$, and inverted neutrino mass ordering (IO), i.e., $m^{}_3 < m^{}_1 < m^{}_2$, we have
\begin{eqnarray} \label{eq:aij2}
\begin{pmatrix}
|a^{}_{11}|^2 & |a^{}_{12}|^2 & |a^{}_{13}|^2 \\
|a^{}_{21}|^2 & |a^{}_{22}|^2 & |a^{}_{23}|^2 \\
|a^{}_{31}|^2 & |a^{}_{32}|^2 & |a^{}_{33}|^2
\end{pmatrix} = \begin{pmatrix}
0.032 & 0.203 & 0.015 \\
0.203 & 0.041 & 0.007 \\
0.015 & 0.007 & 0.229
\end{pmatrix} \; .
\end{eqnarray}
It is worthwhile to notice that $|a_{12}^{}|^2$ (or, equivalently, $|a^{}_{21}|^2$) and $|a_{33}^{}|^2$ are much larger than the others, and they have a more important impact on the statistical determination of neutrino masses. On the other hand, for the six threshold locations $\omega_{ij}^{}$,
it is instructive to study their deviations from $\epsilon_{\rm eg}^{}/2$, i.e., the would-be threshold for massless neutrinos. According to the latest global-fit results~\cite{Gonzalez-Garcia:2014bfa}, we have the best-fit values of two neutrino mass-squared differences $\Delta m_{21}^2 \equiv m^2_2 - m^2_1 = 7.50 \times 10^{-5}_{}~\text{eV}^2$ and $\Delta m_{31}^2 \equiv m^2_3 - m^2_1 = 2.457 \times 10^{-3}_{}~\text{eV}^2$ ($\Delta m_{23}^2 \equiv m^2_2 - m^2_3 = 2.449 \times 10^{-3}_{}~\text{eV}^2$) for NO (IO). Given $\epsilon_{\rm eg}^{} = 2.14349~{\rm eV}$, we plot the deviation $\Delta \omega_{ij}^{} \equiv \epsilon_{\rm eg}^{}/2 - \omega_{ij}^{} = (m^{}_i + m^{}_j)^2/(2\epsilon^{}_{\rm eg})$ as a function of the lightest neutrino mass $m_0^{}$ in Fig.~\ref{fg:dev} for both NO (left) and IO (right) cases. As one can observe from Fig.~\ref{fg:dev}, a hierarchical structure $\Delta \omega^{}_{22}, \Delta \omega^{}_{12}, \Delta \omega^{}_{11} \ll \Delta \omega^{}_{23}, \Delta \omega^{}_{13} \ll \Delta \omega^{}_{33}$ among three groups of deviations in the NO case is obtained, due to the fact that $\Delta m_{21}^2 \ll |\Delta m_{31}^2|$. In the IO case, the hierarchical structure also exists for the same reason but in the opposite order. Moreover, every two groups are separated by a energy gap of $10^{-3}~\text{eV}$, and within each group the separation is around $10^{-5}~\text{eV}$. Thus, in order to resolve all the six thresholds, a trigger laser frequency with a precision better than $10^{-5}~\text{eV}$ is required.

\begin{figure}[t]
\centering
\includegraphics[scale=0.65]{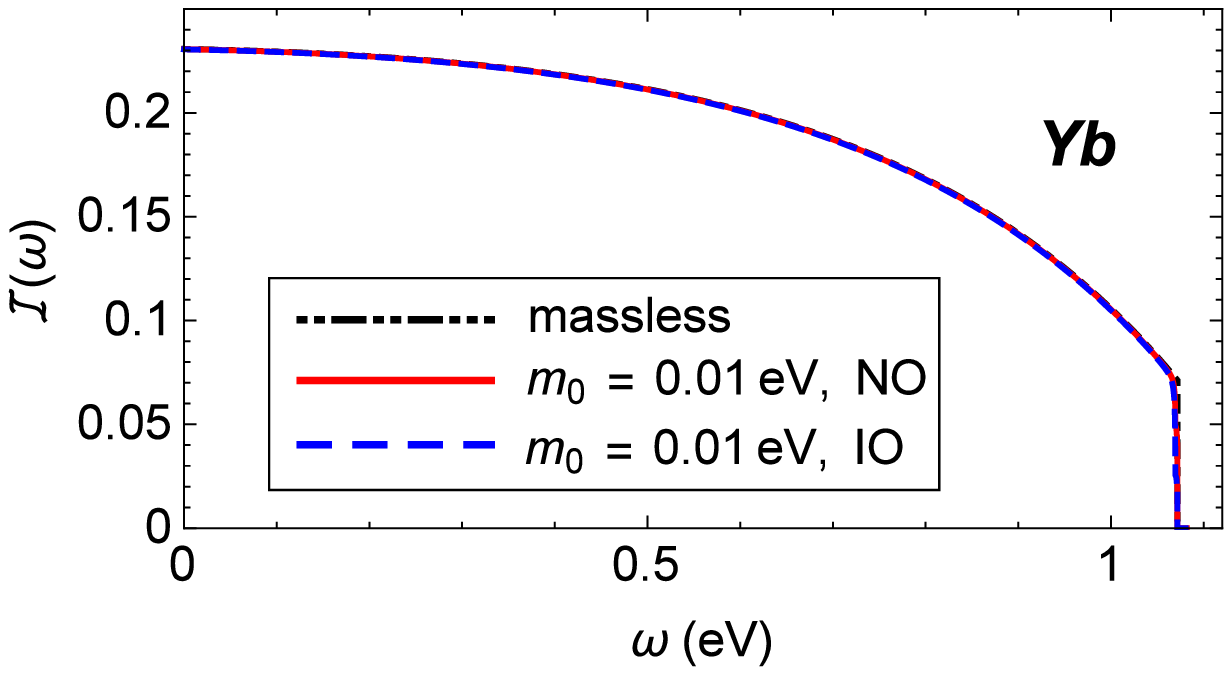} \\
\includegraphics[scale=0.65]{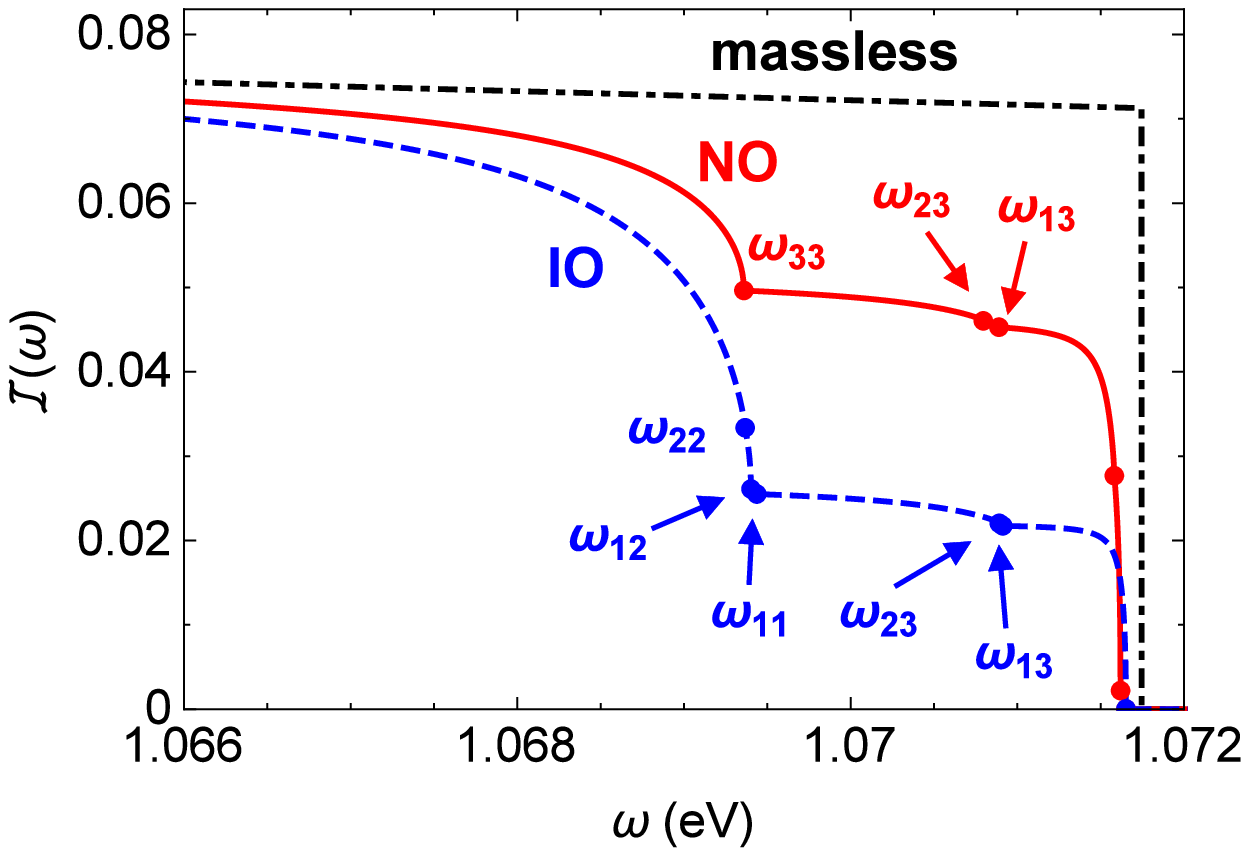}~
\includegraphics[scale=0.65]{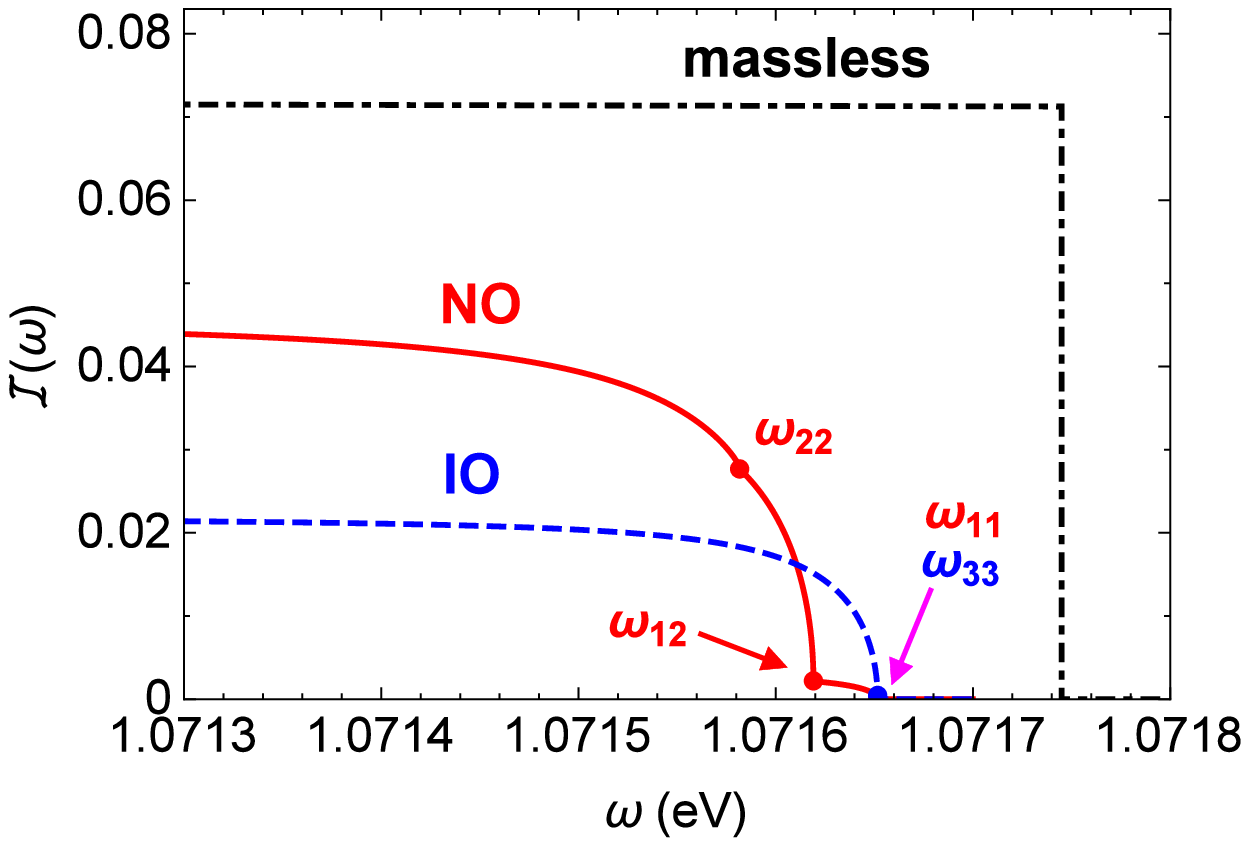}
\caption{Illustration of the spectrum function $\mathcal{I}(\omega)$, where the best-fit values of neutrino mixing parameters from Ref.~\cite{Gonzalez-Garcia:2014bfa} and the values of $\epsilon^{}_{\rm eg}$ and $\epsilon^{}_{\rm vg}$ in Eq.~(\ref{eq:atom_paras}) are used. The black dot-dashed curve denotes the case where all three neutrinos are massless, while red solid and blue dashed curves correspond to the scenarios with $m_0^{} = 0.01~\text{eV}$ in NO and IO, respectively. The plot in the upper panel shows the overall behavior of $\mathcal{I}(\omega)$, with its detailed threshold structures presented in the lower panel. Threshold locations are also indicated by red and blue dots for NO and IO, respectively.}
\label{fg:Iomega_all}
\end{figure}

Now we present some analytical results on the spectrum function $\mathcal{I}(\omega)$, as so far a complete discussion on it is still lacking in the literature. We begin with the case of three massless neutrinos. Although this case is not realistic, it will be helpful for us to see clearly the main difference between massive and massless neutrinos. In this case, only one threshold $\epsilon_{\rm eg}^{}/2 = 1.071745~{\rm eV}$ appears. Near this threshold, we have $\omega \approx \epsilon^{}_{\rm eg}/2$ or $x \equiv \epsilon_{\rm eg}^{}/2 - \omega \ll \epsilon^{}_{\rm eg}$. In consideration of $\epsilon_{\rm eg}^{} \approx \epsilon_{\rm vg}^{}$ as given in Eq.~(\ref{eq:atom_paras}), we can expand the spectrum function $\mathcal{I}(\omega)$ in terms of $x/\epsilon^{}_{\rm eg}$ and obtain
\begin{eqnarray}
\left.\mathcal{I}(\omega)\right|^{}_{m^{}_i = 0} = \frac{1}{12} + \frac{4}{3} \left(\frac{x}{\epsilon_{\rm eg}^{}}\right) + \mathcal{O}\left[ \left( \frac{x}{\epsilon_{\rm eg}^{}} \right)^2 \right] \; .
\end{eqnarray}
Thus, around the threshold $\epsilon_{\rm eg}^{}/2$, the spectrum function $\mathcal{I}(\omega)$ jumps from $0$ to about $1/12$. On the other hand, far away from the threshold, especially near $\omega = 0$, we obtain
\begin{eqnarray}
\left.\mathcal{I}(\omega)\right|^{}_{m^{}_i = 0} = \frac{1}{4} - \frac{11}{12} \left( \frac{\omega}{\epsilon_{\rm eg}^{}} \right)^2_{} + \mathcal{O}\left[\left( \frac{\omega}{\epsilon_{\rm eg}^{}} \right)^3_{}\right] \; .
\end{eqnarray}
These observations agree well with the actual numerical calculation of $\left.\mathcal{I}(\omega)\right|^{}_{m^{}_i = 0}$ given in Fig.~\ref{fg:Iomega_all} (black dot-dashed curve), where the atomic parameters $\epsilon_{\rm eg}^{} = 2.14349~\text{eV}$ and $\epsilon_{\rm vg}^{} = 2.23072~\text{eV}$ are input. Note that the whole spectrum and the partial spectrum around the threshold have been shown in the upper and lower plots of Fig.~\ref{fg:Iomega_all}, respectively.
%%%%%%%%%%%%%%%%%%%%%%%%%%%%%%% Fig. 3 %%%%%%%%%%%%%%%%%%%%%%%%%%%%%%%%%%%%%
%%%%%%%%%%%%%%%%%%%%%%%%%%%%%%%%%%%%%%%%%%%%%%%%%%%%%%%%%%%%%%%

In contrast, massive neutrinos lead to more than one kink structure in the spectrum function. We focus on a single threshold contribution $\mathcal{I}_{ij}^{}(\omega) = \Delta_{ij}^{}(\omega) |a_{ij}^{}|^2 I_{ij}^{}(\omega) \Theta(\omega_{ij}^{} - \omega)/ (\omega-\epsilon_{\rm vg}^{})^2_{}$, since the contributions from different thresholds are independent and will be simply added up to give the full spectrum $\mathcal{I}_{}^{}(\omega) = \sum^{}_{i, j} \mathcal{I}_{ij}^{}(\omega)$. For simplicity, we restrict to the case of $i = j$, namely, $\mathcal{I}_{ii}^{}(\omega)$, and the other cases can be similarly discussed. Near the threshold, namely, $x^\prime \equiv \omega^{}_{ii} - \omega \ll \epsilon^{}_{\rm eg}$, we distinguish two different regions characterized by the relative size of two small parameters $x^\prime/\epsilon^{}_{\rm eg}$ and $2 m_i^2/\epsilon_{\rm eg}^2$:
\begin{itemize}
\item For $x^\prime/\epsilon^{}_{\rm eg} \ll 2 m_i^2/\epsilon_{\rm eg}^2 \ll 1$, we can expand $\mathcal{I}_{ii}^{}(\omega)$ with respect to $2 m_i^2/\epsilon_{\rm eg}^2$ and the ratio of $x^\prime/\epsilon^{}_{\rm eg}$ to $2 m_i^2/\epsilon_{\rm eg}^2$, and arrive at
\begin{eqnarray}\label{eq:expansion_very_near_thres_1}
\mathcal{I}_{ii}^{}(\omega) \approx \frac{|a_{ii}^{}|^2_{}}{6} \sqrt{\frac{\epsilon_{\rm eg}^{}x^\prime_{}}{2 m_i^2}} \; ,
\end{eqnarray}
where the high-order terms have been safely omitted. Thus, for the decreasing frequency $\omega$ just below the threshold, the spectrum function increases as $\mathcal{I}_{ii}^{}(\omega) \propto \sqrt{x^\prime} = \sqrt{\omega^{}_{ii} - \omega}$ from the vanishing value exactly at the threshold. In the case of massless neutrinos, we have already seen a sharp jump from $0$ to about $1/12$.

\item For $2 m_i^2/\epsilon_{\rm eg}^2 \ll x^\prime/\epsilon^{}_{\rm eg} \ll 1$, the spectrum function $\mathcal{I}_{ii}^{}(\omega)$ can be expanded in terms of $2 m_i^2/\epsilon_{\rm eg}^2$ and the ratio of $2 m_i^2/\epsilon_{\rm eg}^2$ to $x^\prime/\epsilon^{}_{\rm eg}$. At the leading order, we have
\begin{eqnarray} \label{eq:expansion_very_near_thres}
\mathcal{I}_{ii}^{}(\omega) \approx |a_{ii}^{}|^2_{} \left( \frac{1}{9} + \frac{16}{9} \frac{x^\prime}{\epsilon^{}_{\rm eg}} -\frac{1}{3} \frac{m_i^4}{x^{\prime 2}_{} \epsilon_{\rm eg}^2}\right) \; .
\end{eqnarray}
It is evident from Eq.~(\ref{eq:expansion_very_near_thres}) that for a frequency $\omega$ not too far away from the threshold, the spectrum function $\mathcal{I}_{ii}^{}(\omega)$ would reach a plateau of $|a_{ii}^{}|^2/9$. One can verify that a similar conclusion is valid for the $i \neq j$ case and the plateau becomes $|a_{ij}^{}|^2/9$. Noticing the identity $\sum_{ij}^{} |a_{ij}^{}|^2 = 3/4$, we then obtain an asymptotic value of $1/12$ for $\mathcal{I}(\omega)$ when including all the threshold contributions.
\end{itemize}
Therefore, we can understand the main effect of nonzero neutrino masses, i.e., replacing the sharp jump in the massless case with fine kink structures. Such an observation agrees perfectly with the numerical results shown in Fig.~\ref{fg:Iomega_all}, where the fine structure can be clearly seen in the two plots in the lower panel.

Let us comment on the prominence of various kinks in the spectrum function, as shown in Fig.~\ref{fg:Iomega_all}. Since each threshold contribution is proportional to the factor $|a_{ij}^{}|^2$, we can conclude that the prominent kinks appear at the thresholds of $\omega_{12}^{}$ and $\omega_{33}^{}$ for both NO and IO according to Eq.~(\ref{eq:aij2}), although the order of their appearance is reversed, namely, $\omega_{12}^{} > \omega_{33}^{} (\omega_{12}^{} < \omega_{33}^{})$ for NO (IO).

\section{Statistical Determination of Neutrino Masses}

Now we discuss the conditions for a statistical determination of neutrino masses, and try to improve the analysis performed in Ref.~\cite{Song:2015xaa}. In order to probe absolute neutrino masses in
a real RENP experiment, we expect that a two-step scanning strategy, i.e., a rough scan followed by a fine one, may be adopted. In the first rough scan one locates the trigger laser frequency at several places so as to obtain a fair knowledge about the absolute neutrino mass scale, while it is in the later fine scan that we determine the absolute neutrino mass scale accurately. In the present work, we focus on the fine scan and aim at answering the question how to set the location of the trigger laser frequency so that a minimal detection time is required to reach a given accuracy of the measurement of neutrino masses. Later in this section we will comment on the required accuracy for the rough scan. In addition, we assume no background processes in this study.\footnote{Although the allowed two-photon transition $|{\rm e}\rangle \to |{\rm g}\rangle + |\gamma\rangle + |\gamma\rangle$ can be an intrinsic background, there exist possible ways of circumventing it, as discussed in Ref.~\cite{Fukumi:2012rn}.}

The same question has recently been studied in Ref.~\cite{Song:2015xaa}, in which the location of the trigger laser frequency is set to be around the \emph{first} threshold close to $\epsilon_{\rm eg}^{}/2$ for both NO and IO cases. Although such a choice is indeed optimal for the IO case (as we will explain later), it is not for NO, especially for the lightest neutrino mass $m_0^{} \sim 0.01~\text{eV}$. To see this point clearly, we consider the discrimination of the $m_0^{} = 0.01~\text{eV}$ and $m_0^{} = 0.011~\text{eV}$ cases in NO, namely, a $10\%$ accuracy of determination for a true value of $m_0^{} = 0.01~\text{eV}$.
%%%%%%%%%%%%%%%%%%%%%%%%%%%%%%%% Fig. 4 %%%%%%%%%%%%%%%%%%%%%%%%%
\begin{figure}[t]
\centering
\includegraphics[scale=0.75]{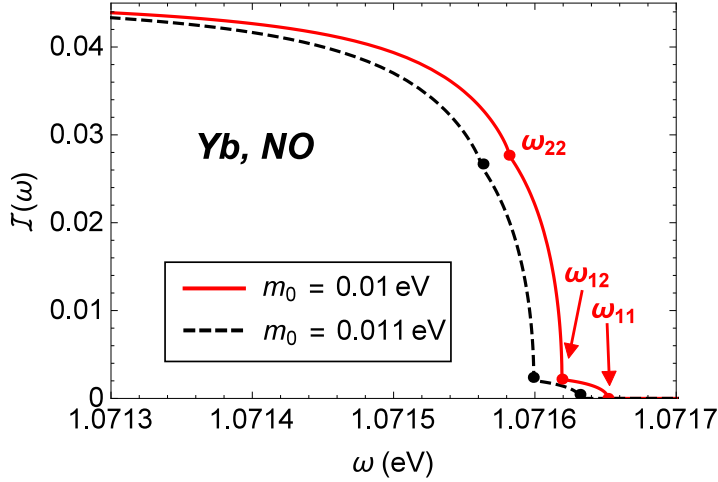}
\caption{First three threshold structures of spectrum function $\mathcal{I}(\omega)$ for $m_0^{} = 0.01~\text{eV}$ (red solid) and $m_0^{} = 0.011~\text{eV}$ (black dashed) in the NO case.}
\label{fg:detailed_threshold}
\end{figure}
%%%%%%%%%%%%%%%%%%%%%%%%%%%%%%%%%%%%%%%%%%%%%%%%%%%%%%%%%
In Fig.~\ref{fg:detailed_threshold} we show the first three threshold structures of spectrum function $\mathcal{I}(\omega)$ for $m_0^{} = 0.01~\text{eV}$ (red solid) and $m_0^{} = 0.011~\text{eV}$ (black dashed) in NO. As one can see, if one locates the trigger frequency at the first threshold $\omega_{11}^{}$ for the $m_0^{} = 0.011~\text{eV}$ case, although zero events are obtained for $m_0^{} = 0.011~\text{eV}$, the rate of obtaining events for $m_0^{} = 0.01~\text{eV}$ is also very small, i.e., $\mathcal{I}(\omega^{}_{11}) \sim 0.001$. This explains why there exists a sharp increase in the detection time when decreasing $m_0^{}$ towards $\sim 0.01~\text{eV}$ in Fig.~2 of Ref.~\cite{Song:2015xaa} (also reproduced as the black dashed curve in Fig.~\ref{fg:Nnorm_comp_sigma_m}). Apparently, Fig.~\ref{fg:detailed_threshold} suggests other better places to locate the trigger laser frequency, e.g., the prominent $\omega_{12}^{}$ threshold.

To systematically determine the optimized trigger laser frequency, we first introduce the procedure of our statistical analysis. For a given trigger laser frequency $\omega$, a true value of $m_0^{}$ and a detection time, we define the exclusion probability of the $(1+\sigma_{m})m_0^{}$ case (corresponding to an accuracy $\sigma_{m}$) by assuming Poisson statistics, i.e., $P_{\text{ex}}^{} = 1 - \text{Poiss}(N_{m_0}^{}, N_{(1+\sigma_{m})m_0}^{})$, where $P_{\text{ex}}^{}$ is the exclusion probability, $N_{m_0}^{}$ and $N_{(1+\sigma_{m})m_0}^{}$ are the numbers of expected events in the two competing cases, respectively, and $\text{Poiss}(\mu,\lambda)$ is the probability of observing $\lambda$ events with an expected value of $\mu$ events in Poisson statistics. With this definition, if requiring a $3\sigma$ exclusion, i.e., $P_{\text{ex}}^{} = 0.9973$, we obtain the detection time for a given trigger laser frequency and a true value of $m_0^{}$. Following Ref.~\cite{Song:2015xaa}, we normalize the detection time by nominal values of the target volume and the number density of atoms in the target
\begin{eqnarray}
N_{\text{norm}}^{} = \left( \frac{T}{\text{sec}} \right) \left( \frac{V}{10^2_{}~\text{cm}^3_{}} \right) \left( \frac{n}{10^{21}_{}~\text{cm}^{-3}_{}} \right)^3 \; .
\end{eqnarray}
Furthermore, to avoid any confusion, we also distinguish the threshold locations in the $m_0^{}$ and $(1+\sigma_m^{})m_0^{}$ cases by using $\omega_{ij}^{}$ and $\omega_{ij}^\prime$ for the former and latter, respectively.
%%%%%%%%%%%%%%%%%%%%%%%%%%%%%%%%%% Fig. 5 %%%%%%%%%%%%%%%%%%%%%%%%%%
\begin{figure}[t]
\centering
\includegraphics[scale=0.67]{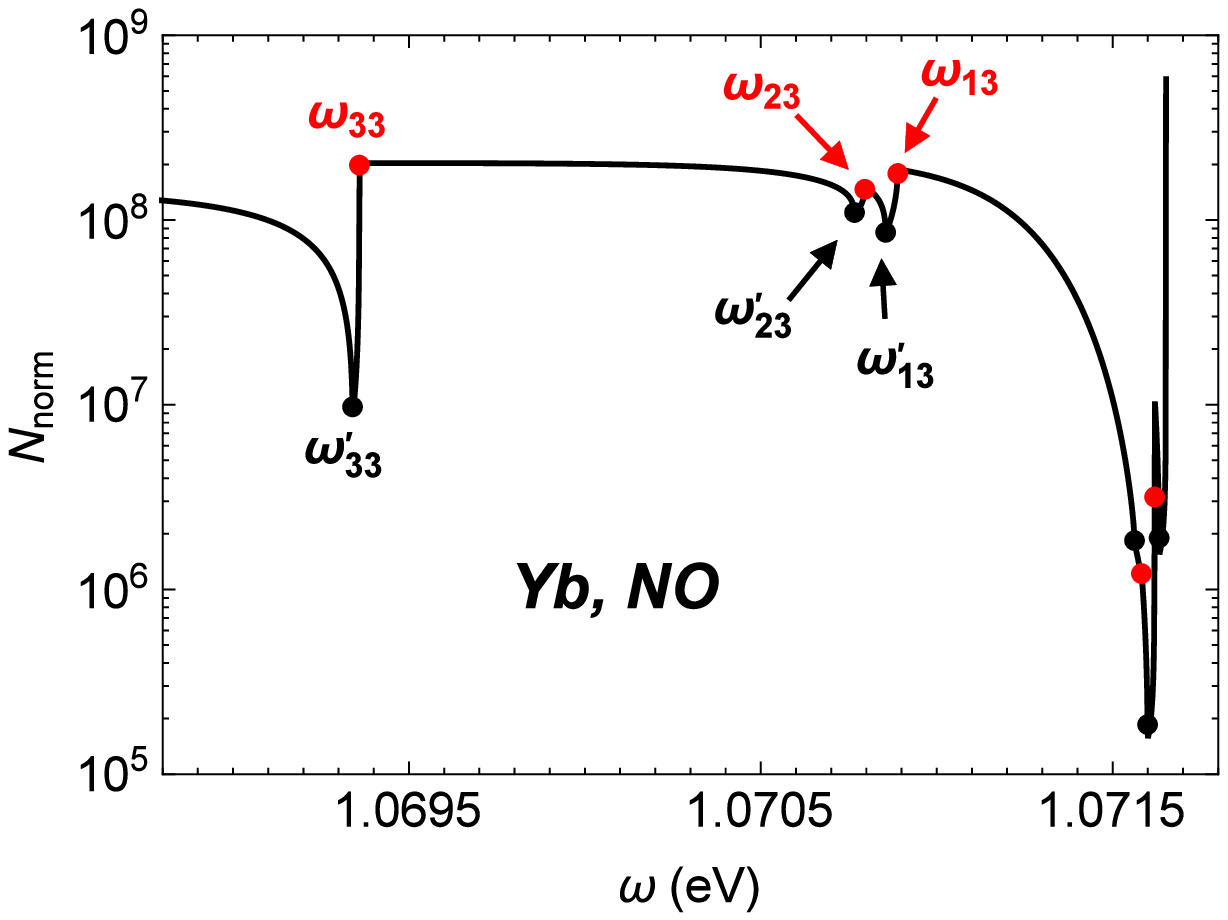}~
\includegraphics[scale=0.67]{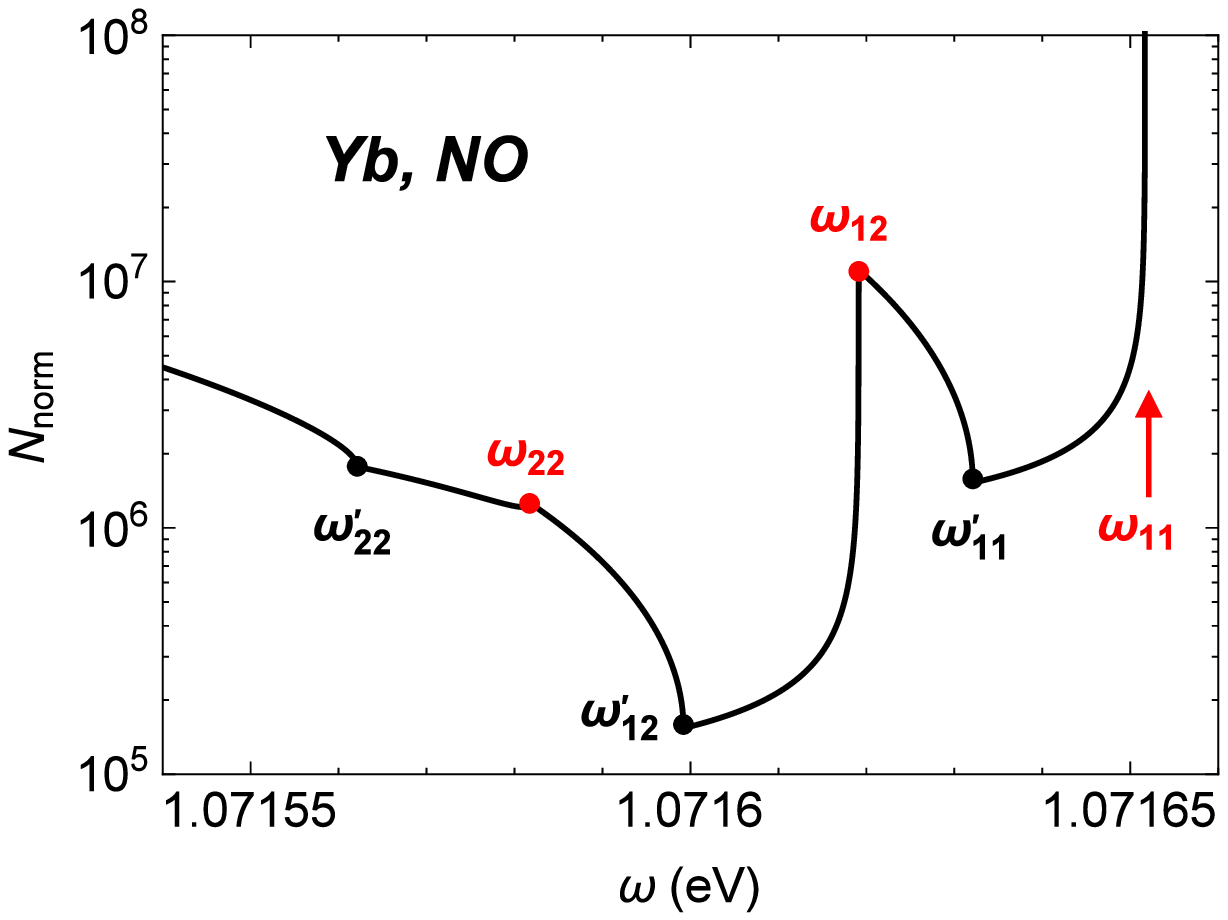}
\vspace{-0.0cm}
\caption{The normalized detection time $N_{\text{norm}}^{}$ as a function of the trigger laser frequency $\omega$, where $m_0^{} = 0.01~\text{eV}$ is taken in the NO case and a $3\sigma$ exclusion for a $10\%$ level of accuracy is assumed. Black dots indicate the threshold locations in the $m_0^{} = 0.011~\text{eV}$ case, while red dots in the $m_0^{} = 0.01~\text{eV}$ case. The left plot provides the overall picture for a large range of $\omega$, while the detailed structures near the first three thresholds are given in the right plot.}
\label{fg:scan}
\end{figure}
%%%%%%%%%%%%%%%%%%%%%%%%%%%%%%%%%%%%%%%%%%%%%%%%%%%%%%%%%%%%%%%%

In Fig.~\ref{fg:scan} we show $N_{\text{norm}}^{}$ as a function of the trigger laser frequency $\omega$. In the calculations, we take $m_0^{} = 0.01~\text{eV}$ in the NO case and assume a $3\sigma$ exclusion for a $10\%$ level of accuracy (i.e., $\sigma_m^{} = 10\%$). Black dots indicate the threshold locations in the $m_0^{} = 0.011~\text{eV}$ case, while red dots in the $m_0^{} = 0.01~\text{eV}$ case. The left plot provides the overall picture for a large range of $\omega$, while the detailed structures near the first three thresholds are given in the right plot. Then, it becomes clear that locating the trigger laser frequency at $\omega_{12}^\prime$ yields the least detection time, which is about one order of magnitude smaller than that at $\omega_{11}^\prime$, a trigger laser frequency used in Ref.~\cite{Song:2015xaa}. In fact, it is unnecessary to set $\omega$ precisely at $\omega_{12}^\prime$, as there exists a range of $\omega$ near $\omega_{12}^\prime$ that can also yield a decent detection time. Such a loose requirement actually provides a stopping criterion for the rough scan, namely, one may terminate the rough scan, as long as the location of $\omega_{12}^\prime$ is known to an accuracy that is adequate for locating the trigger laser in the fine scan. A similar scan for the IO case has also been performed (but not shown here). As the first threshold $\omega_{33}^{\prime}$ in the IO case is already quite prominent, as can be seen from Fig.~\ref{fg:Iomega_all}, we find that $\omega_{33}^{\prime}$ is indeed optimal for the IO case. Hence, we suggest using the thresholds of $\omega_{12}^{\prime}$ and $\omega_{33}^{\prime}$ for NO and IO, respectively.

Since the results given in Fig.~2 of Ref.~\cite{Song:2015xaa} remain appropriate for the IO case, we only update the statistical analysis of the NO case by locating the trigger laser frequency at $\omega_{12}^{\prime}$. In Fig.~\ref{fg:Nnorm_comp_sigma_m}, we show $N_{\text{norm}}^{}$ as a function of $m_0^{}$, assuming a $3\sigma$ exclusion for three levels of accuracy, i.e., $\sigma_m = 1\%$ (red), $10\%$ (black) and $50\%$ (blue). Dashed curves are the results presented in Fig.~2 of Ref.~\cite{Song:2015xaa}, while the solid ones are our improved results. It can be seen that for $m_0^{} \sim 0.01~\text{eV}$, adopting the threshold of $\omega_{12}^{\prime}$ can reduce the detection time by around one order of magnitude if a $3\sigma$ exclusion for an accuracy of better than $10\%$ is required. For a much smaller true value of $m_0^{}$, it seems to be optimal by still using $\omega_{11}^{\prime}$.\footnote{The plateau structure of the dashed curves in Fig.~\ref{fg:Nnorm_comp_sigma_m} can be understood by adopting the analytical expansion in Eq.~(\ref{eq:expansion_very_near_thres_1}). For smaller values of $m^{}_0$, the difference between two spectrum functions at $\omega_{11}^\prime$ is given by $|a_{11}^{}|^2_{}\sqrt{\sigma_m^2 + 2\sigma_m^{}}/6$ at the leading order, so it is actually independent of the absolute neutrino masses.} However, one should keep in mind that in reality the uncertainty of the trigger laser frequency may be the major issue in that range of $m_0^{}$, resulting in almost no discriminating power on neutrino masses, no matter which threshold is adopted. More discussions on the effect of the uncertainty of the trigger laser frequency can be found in Ref.~\cite{Song:2015xaa}.
%%%%%%%%%%%%%%%%%%%%%%%%%%%%%%% Fig. 6 %%%%%%%%%%%%%%%%%%%%%%%%%%
\begin{figure}[t]
\centering
\includegraphics[scale=0.7]{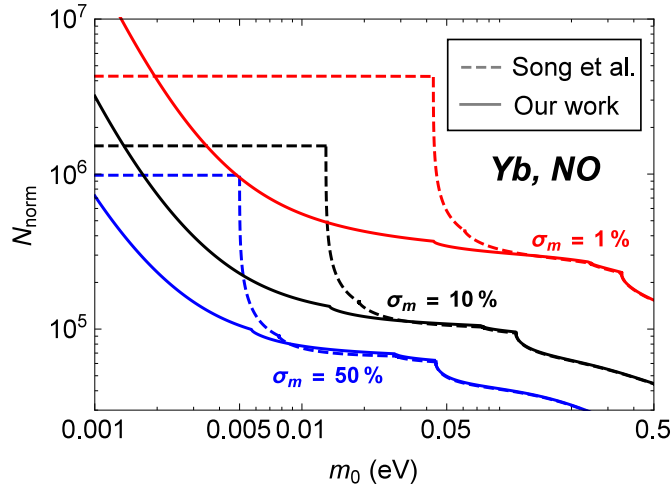}
\vspace{-0.0cm}
\caption{The normalized detection time $N_{\text{norm}}^{}$ as a function of the true value of $m_0^{}$, where a $3\sigma$ exclusion for the accuracy of $\sigma_m = 1\%$ (red), $10\%$ (black) and $50\%$ (blue) is assumed. Dashed curves are the results in Fig.~2 of Ref.~\cite{Song:2015xaa}, while solid ones stand for our results for which the threshold of $\omega_{12}^{\prime}$ has been adopted.}
\label{fg:Nnorm_comp_sigma_m}
\end{figure}
%%%%%%%%%%%%%%%%%%%%%%%%%%%%%%%%%%%%%%%%%%%%%%%%%%%%

\section{Conclusions}

One of the burning questions in neutrino physics at present is to determine the absolute scale $m_0^{}$ of neutrino masses. Although there exist several traditional approaches of measuring it, such as studying $\beta$ and $0\nu\beta\beta$ decays in nuclear physics and inferring it from the cosmological data, a sensitivity at the level of $m_0^{} \sim 0.01~\text{eV}$ is very difficult to achieve. Thus, it is imperative to explore other possible ways.

In this work we concentrate on the novel idea of measuring neutrino masses in atomic physics by observing an atomic transition accompanied by a radiative emission of a neutrino pair and a photon~\cite{RENP_idea}. First, we give an analytical description of the photon spectrum in such a RENP process. The fine structures around different thresholds can be well understood. Then, we improve the statistical determination of absolute neutrino masses recently performed in Ref.~\cite{Song:2015xaa}. In particular, in the NO case, we find that the best choice of a trigger laser frequency is close to $\omega^{}_{12}$ instead of $\omega^{}_{11}$. As a consequence, the previously required detection time can be reduced by one order of magnitude for the case of a $3\sigma$ determination of $m_0^{} \sim 0.01~\text{eV}$ with an accuracy better than $10\%$. Such an improvement is very encouraging for the future investigations on measuring neutrino masses in atomic processes.

Finally, it is worthwhile to emphasize again the importance of probing intrinsic properties of neutrinos in atomic physics. As the characteristic energy scale in atomic processes is comparable to neutrino masses, the determination of absolute neutrino masses and the discrimination between Dirac and Majorana neutrinos in the nonrelativistic region seem to be quite promising. In this sense, atomic physics offers us an ideal and unique tool to resolve the puzzles on neutrinos. However, further great theoretical and experimental efforts in this direction are needed to accomplish a real experiment and achieve its main physics goals.

\hspace{0.2cm}
\begin{flushleft}
{\bf Acknowledgements}
\end{flushleft}
The authors are grateful to Professor M. Yoshimura for helpful correspondence and Guo-yuan Huang for valuable discussions. This work was supported in part by the National Recruitment Program for Young Professionals and by the CAS Center for Excellence in Particle Physics (CCEPP).

\end{document}